\pdfoutput=1
\documentclass[twocolumn]{autart}    

\usepackage{amssymb}
\usepackage{amsmath}
\usepackage{breqn}  
\usepackage{graphicx}
\usepackage{hyperref}

\usepackage{algorithm}% <=================http://ctan.org/pkg/algorithms
\usepackage{algpseudocode}% <========== http://ctan.org/pkg/algorithmicx

\begin{document}

\begin{frontmatter}

\title{Nonlinear system identification with regularized\\ Tensor Network B-splines} % Title, preferably not more 

\author[Delft]{Ridvan Karagoz}\ead{r.karagoz@hotmail.com},    % Add the 
\author[Delft]{Kim Batselier}\ead{K.Batselier@tudelft.nl},               % e-mail address % (ead) as shown
 % Please supply                                              

\address[Delft]{Delft Center for Systems and Control, Delft University of Technology, The Netherlands}        % here.

                             % keyword list or with the 

\begin{abstract}                          % Abstract of not more than 200 words.
This article introduces the Tensor Network B-spline model for the regularized identification of nonlinear systems using a nonlinear autoregressive exogenous (NARX) approach. Tensor network theory is used to alleviate the curse of dimensionality of multivariate B-splines by representing the high-dimensional weight tensor as a low-rank approximation. An iterative algorithm based on the alternating linear scheme is developed to directly estimate the low-rank tensor network approximation, removing the need to ever explicitly construct the exponentially large weight tensor. This reduces the computational and storage complexity significantly, allowing the identification of NARX systems with a large number of inputs and lags. The proposed algorithm is numerically stable, robust to noise, guaranteed to monotonically converge, and allows the straightforward incorporation of regularization. The TNBS-NARX model is validated through the identification of the cascaded watertank benchmark nonlinear system, on which it achieves state-of-the-art performance while identifying a 16-dimensional B-spline surface in 4 seconds on a standard desktop computer. An open-source MATLAB implementation is available on GitHub.
\end{abstract}

\end{frontmatter}

\section{Introduction}

B-splines are basis functions for the spline function space \cite{de1976splines}, making them an attractive choice for approximating smooth continuous functions. For this reason, B-splines have had numerous applications in system identification \cite{lightbody1997neural,yiu2001nonlinear,lin2007online,dos2009nonlinear,mirea2014dynamic,folgheraiter2016combined,csurcsia2014nonparametric} and control \cite{cong2000improved,reay2003cmac,hong2012system,zhang2017adaptive}.  The generalization of B-splines to multiple dimensions is done through tensor products of their univariate basis functions. The number of basis functions and weights that define a multivariate B-spline surface, therefore, increase exponentially with the number of dimensions, i.e. B-splines suffer from the curse of dimensionality. Previous attempts to avoid this limitation include strategies such as dimensionality reduction, ANOVA decompositions and hierarchical structures \cite{brown1995high}. The most effective method, i.e. hierarchical B-splines, relies on sparse grids \cite{zenger1991sparse} and reduces the storage complexity from $\mathcal{O}(k^{d})$ to $\mathcal{O}(k \, \text{log}(k)^{d-1})$ \cite{garcke2006sparse}. This is still exponential in the number of dimensions $d$. 
A recently emerging way to alleviate the curse of dimensionality is through the concept of tensor networks. Originally developed in the context of quantum physics, tensor networks efficiently represent high-dimensional tensors as a set of sparsely interconnected low-order tensors \cite{cichocki2016tensor}. Combined with tensor algebra, tensor network structures can greatly decrease the computational complexity of many applications \cite{batselier2017kalman,batselier2017tensor,batselier2018tensor}. Due to their multilinear nature, multivariate B-splines easily admit a tensor network representation, which we call the Tensor Network B-splines (TNBS) model. Algorithms for optimization in the tensor network format make it possible to fit multivariate B-spline surfaces onto high-dimensional data by directly finding a low-rank tensor network approximation of the weight tensor, thereby overcoming the curse of dimensionality. This broadens the applicability of multivariate B-splines to high-dimensional problems that often occur in system identification and control.
One particularly well-suited application of Tensor Network B-splines is black-box nonlinear system identification. The Nonlinear Autoregressive eXogenous (NARX) model \cite{leontaritis1985input} is able to represent a wide range of nonlinear systems and is useful when knowledge about the model structure of the system is limited. For the single-input–single-output (SISO) case, the discrete-time NARX model is expressed by the following nonlinear difference equation:
\begin{equation}\label{system}
     y_n = f(y_{n-1}, y_{n-2}, \ldots, u_{n}, u_{n-1}, u_{n-2}, \ldots) + \varepsilon_n.
\end{equation}
The function $f$ is an unknown nonlinear mapping and $u_n$ and $y_{n}$ are the input and output samples at time step $n$. The error $\varepsilon_t$ is assumed to be Gaussian white noise. The most common models used in approximating $f$ are polynomials or neural networks \cite{billings2013nonlinear}. The applicability of polynomial NARX is, however, often limited to weakly nonlinear systems due to computational complexity. Neural networks, on the other hand, require a lot of data to generalize well and can be time consuming to train. Under the reasonable assumption that $f$ is sufficiently smooth, the Tensor Network B-splines model is a suitable candidate to approximate the function from observed input and output data. The contributions of this paper are:
\begin{itemize}
    \item Introduce the Tensor Network B-splines model.
    \item Present a regularized TNBS-NARX system identification algorithm.
\end{itemize}
The paper is structured as follows. Section 2 introduces relevant tensor and B-spline theory. Section 3 presents the TNBS model, the regularization technique and the NARX identification algorithm. Section 4 validates the TNBS-NARX approach through numerical experiments on a synthetic and a benchmark dataset. Section 5 concludes this paper and lists some recommendations.

\section{Preliminaries}

This section provides the basic terminology and definitions for tensors and tensor decompositions, followed by an introduction to B-splines. Most of the introduced tensor network definitions are based on \cite{kolda2009tensor,penrose1971applications,cichocki2014era,oseledets2011tensor}. A comprehensive treatment of B-splines is given in the book by de Boor \cite{practical}.
\subsection{Tensor basics}

A tensor is a multidimensional array of real numerical values, e.g. $\mathcal{A} \in \mathbb{R}^{k_1\times k_2\times \cdots \times k_d}$. Tensors can thus be considered generalizations of vectors and matrices. The order $d$ of the tensor is the number of dimensions of the array. Unless stated otherwise, subscript indices indicate a single element of a tensor, e.g. $a = \mathcal{A}_{i_1, i_2, \ldots, i_d}$. The size of each dimension is indicated by $k_p, p \in \{1,2, \ldots, d\} $, such that $i_p \in \{1,2, \ldots, k_p\}$. In this paper, scalars are denoted by lowercase letters ($a$), vectors are denoted by bold lowercase letters ($\boldsymbol{a}$), matrices are denoted by bold uppercase letters ($\boldsymbol{A}$) and higher-order tensors are denoted by calligraphic letters ({$\mathcal{A}$}). \begin{figure}[h]
    \begin{center}
    \includegraphics[scale=0.2]{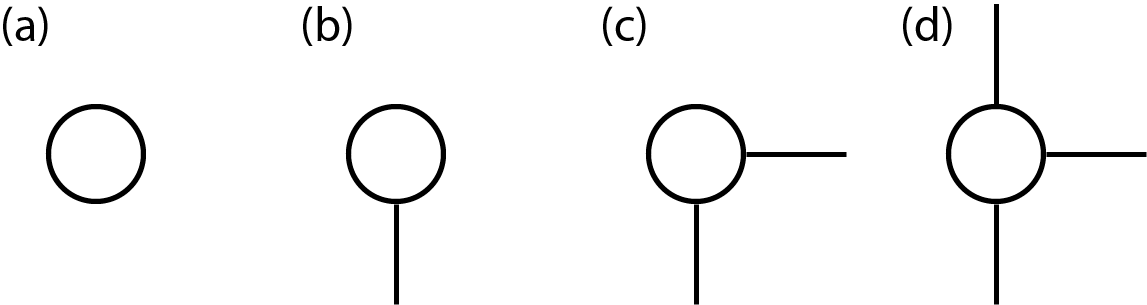}      
    \caption{Graphical notation of a (a) scalar, (b) vector, (c) matrix and (d) third-order tensor.}
    \label{fig:tenz}                                
    \end{center}                                 
\end{figure}
\begin{figure}[b]
    \begin{center}
    \includegraphics[scale=0.2]{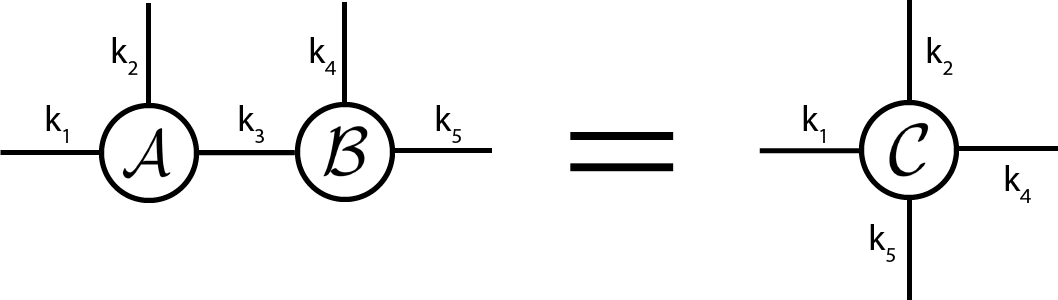}      
    \caption{Tensor contraction in graphical notation.}  
    \label{fig:graphical2}                                
    \end{center}                                 
\end{figure} A convenient way of expressing tensors and their operations is using the graphical notation introduced by Roger Penrose in 1972 \cite{penrose1971applications}. Figure \ref{fig:tenz} shows the representation of a scalar, vector, matrix and third-order tensor using this notation. Every node represents a tensor, the edges represent the indices and the number of edges, therefore, corresponds to its order. The vectorization of a tensor $\mathcal{A} \in \mathbb{R}^{k_1\times k_2  \times \cdots \times k_d}$ is the reordering of its elements into a column vector, denoted by $\boldsymbol{vec}(\mathcal{A}) = \boldsymbol{a} \in \mathbb{R}^{k_1 k_2\cdots k_d}$. The elements of $\boldsymbol{a}$ are denoted as:
\begin{equation*}
\boldsymbol{a}_{i_1 + (i_2-1)k_1 + \ldots + (i_d-1)k_1 k_2 \ldots k_{d-1}} =\mathcal{A}_{i_1, i_2, \ldots, i_d}.
\end{equation*}
A tensor $\mathcal{T} \in \mathbb{R}^{k_1\times k_2  \times \cdots \times k_d}$ is of rank one if it can be decomposed into the outer product of $d$ vectors \mbox{$\boldsymbol{b}^{(p)} \in \mathbb{R}^{k_p}$}, e.g: 
$$\mathcal{T} =\boldsymbol{b}^{(1)} \circ \boldsymbol{b}^{(2)}\circ \cdots \circ \boldsymbol{b}^{(d)},$$
where $\circ$ denotes the outer product operation. The most essential operation in tensor algebra is contraction, which is the summing of elements over equal-sized indices. Given the tensors $\mathcal{A} \in \mathbb{R}^{k_1\times k_2 \times k_3} $ and $\mathcal{B} \in \mathbb{R}^{k_3\times k_4 \times k_5} $, contracting the index $i_3$ results in a tensor $\mathcal{A} \, \times_{3}^{1} \, \mathcal{B} = \mathcal{C} \in \mathbb{R}^{k_1\times k_2 \times k_4 \times k_5}$ whose elements are given by:
\begin{equation}\label{contrac}
    \mathcal{C}_{i_1,i_2,i_4,i_5} = \sum_{i_3} \mathcal{A}_{i_1,i_2,i_3} \hspace{1mm} \mathcal{B}_{i_3,i_4,i_5}.
\end{equation}
Contraction is indicated by the left-associative \mbox{$\binom{m}{n}$-mode} product operator \cite{cichocki2016tensor}, where $n$ and $m$ indicate the position of the indices of the first and second tensor respectively. In the graphical notation, contraction is indicated by connecting corresponding edges, as illustrated for \eqref{contrac} in Figure \ref{fig:graphical2}. An important equation \cite{batselier2017tensor} that relates contraction of a $d$-dimensional tensor with $d$ matrices to a linear operation is the following:
\begin{align}\label{kim}
\boldsymbol{vec}\left(\mathcal{A}  \times_{1}^2 \boldsymbol{C}^{(1)}  \times_{2}^2 \cdots \times_{d}^2 \boldsymbol{C}^{(d)}\right) \nonumber\\
= \left(\boldsymbol{C}^{(d)} \otimes \cdots \otimes \boldsymbol{C}^{(1)} \right) \boldsymbol{vec}(\mathcal{A}),
\end{align}
where $\otimes$ denotes the Kronecker product. The outer product operation is a special case of contraction where the contracted indices have singleton dimensions. The outer product is depicted in the graphical notation by a dashed line connecting two nodes. The inner product between two equal-sized tensors is the sum of their entry-wise products, equivalent to contraction of the tensors over all pairs of indices. Given two tensors $ \mathcal{A} \in \mathbb{R}^{k_1\times k_2 \times k_3} $ and $\mathcal{B} \in \mathbb{R}^{k_1\times k_2 \times k_3}$, their inner product is given by:
\begin{center}
$\left< \mathcal{A},\mathcal{B} \right> = \sum_{i_1,i_2,i_3}\mathcal{A}_{i_1,i_2,i_3}\mathcal{B}_{i_1,i_2,i_3} = \boldsymbol{vec}(\mathcal{A})^T \boldsymbol{vec}(\mathcal{B})$.
\end{center}
The Frobenius norm of a tensor is defined as the square root of the sum of squares of its entries:
\begin{equation*}
\|\mathcal{A}\|_{2} = \sqrt{\left< \mathcal{A},\mathcal{A} \right>}.
\end{equation*}

\subsection{Tensor trains}

\begin{figure}
    \begin{center}
    \includegraphics[scale=0.2]{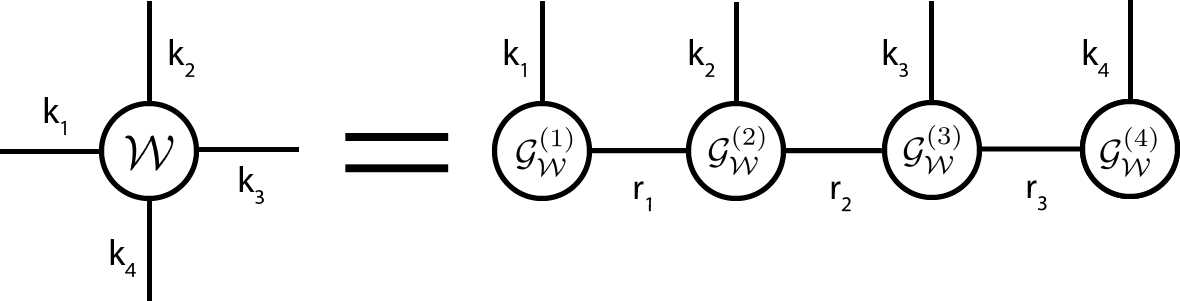}      
    \caption{Graphical notation of the tensor train decomposition for a fourth-order tensor.}  
    \label{fig:TT}                                
    \end{center}                                 
\end{figure}
The tensor train (TT) decomposition is a widely used tensor network format, popular for its low parametric format and the numerical stability of related optimization algorithms \cite{oseledets2011tensor}. A tensor train expresses a tensor $\mathcal{W} \in \mathbb{R}^{k_1\times k_2\times \cdots \times k_d} $ of order $d$ in terms of third-order tensors $\mathcal{G}^{(p)}_{\mathcal{W}} \in \mathbb{R}^{r_{p-1} \times k_p\times r_{p}}$, also known as the TT-cores.
Figure \ref{fig:TT} shows the TT-decomposition of a four-dimensional tensor in graphical notation. The dimensions of the contracted indices, $r_{p}$, are called TT-ranks. The first and last TT-ranks, $r_{0}$ and $r_{d}$, are by definition equal to one. Keeping in mind that the \mbox{$\binom{m}{n}$-mode} product operator is left-associative, the tensor train in Figure \ref{fig:TT} can be expressed as: 
\begin{align}\label{ttw}
\mathcal{W} = \mathcal{G}_{\mathcal{W}}^{(1)} \times_{2}^1 \mathcal{G}_{\mathcal{W}}^{(2)} \times_{3}^1 
\mathcal{G}_{\mathcal{W}}^{(3)} \times_{4}^1 
\mathcal{G}_{\mathcal{W}}^{(4)}.
\end{align}
There exists a set of TT-ranks $r_p = R_{p}$ for which the decomposition is exact. When $r_p < R_{p}$, the tensor train represents an approximation of the original tensor. The lower the TT-ranks, the less accurate the decomposition, but the better the compression. When all $r_{p}$ and dimensions $k_p$ are equal, the storage complexity of the tensor train representation is $\mathcal{O}(kdr^2)$. A TT-decomposition with low TT-ranks can thus significantly reduce the memory footprint of high-dimensional data. For a prescribed set of TT-ranks or a prescribed accuracy, the TT-decomposition of a tensor can be computed with the TT-SVD \cite{oseledets2011tensor} or the TT-Cross \cite{oseledets2010tt} algorithm. An important notion for TT-cores is orthogonality. A TT-core $\mathcal{G}^{(p)}_{\mathcal{W}}$ is left-orthogonal if it can be reshaped into a matrix $\boldsymbol{G}^{(p)} \in \mathbb{R}^{r_{p-1} k_p\times r_{p}}$  for which:
$$\boldsymbol{G}^{(p)T} \boldsymbol{G}^{(p)} = \boldsymbol{I}.$$
Likewise, $\mathcal{G}^{(p)}_{\mathcal{W}}$ is is right-orthogonal if it can be reshaped into a matrix $\boldsymbol{G}^{(p)}\in \mathbb{R}^{r_{p-1}\times k_p r_{p}}$  for which: $$\boldsymbol{G}^{(p)} \boldsymbol{G}^{(p)T} = \boldsymbol{I}.$$
A tensor train is in site-k-mixed-canonical form \cite{schollwock2011density} when for for its TT-cores the following applies:
\begin{align}\label{canon}
\mathcal{G}^{(p)}_{\mathcal{W}} = \left\{
\begin{matrix} 
\text{left-orthogonal,\,} & 1 \leq p \leq k-1 \\
\text{right-orthogonal,\,} & k+1 \leq p \leq d.
\end{matrix}
\right.
\end{align}
For a site-k-mixed-canonical tensor train holds that its norm is contained in the $k$-th TT-core, i.e.:
\begin{equation*}
\|\mathcal{W}\|_{2} = \|\mathcal{G}^{(k)}_{\mathcal{W}}\|_{2} .
\end{equation*}

\subsection{B-splines}
\begin{figure}[t]
    \centering
    \includegraphics[width=8.4cm]{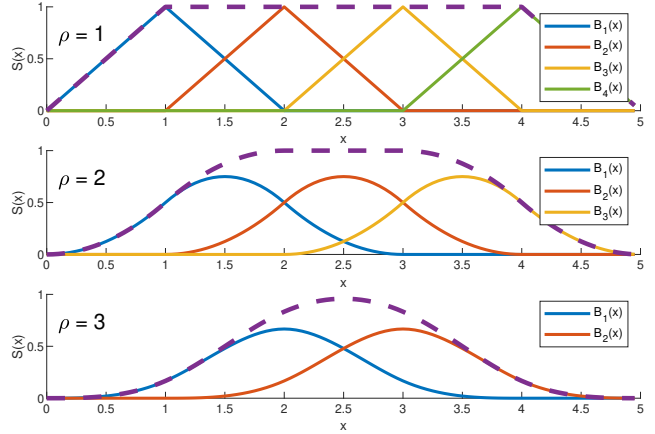}
    \caption{Cardinal B-splines of degrees $1$ to $3$. The dashed purple lines represent the sum of the B-splines.}
    \label{fig:cardinal}
\end{figure}
A univariate spline $S$ is a piecewise polynomial function that maps values from an interval $[a,b]$ to the set of real numbers, e.g. $S: [a,b] \in \mathbb{R} \to \mathbb{R}$. Any spline of degree $\rho$ can be expressed as a unique linear combination of B-splines of the same degree:
\begin{align}\label{basic}
{S(x) =\sum_{i=1}^k B_{i}(x) w_i 
= \boldsymbol{b}^T  \boldsymbol{w}}\\
= \begin{bmatrix}
   B_{1,\rho}(x) & B_{2,\rho}(x) & \cdots & B_{k,\rho}(x)
\end{bmatrix}
\begin{bmatrix}
  w_1 \\ w_2 \\ \vdots \\ w_k
\end{bmatrix}.
\end{align}
The B-spline basis functions $B_{i,\rho}(x)$ are defined by the knot sequence and degree $\rho$, and they are contained in the basis vector $\boldsymbol{b}$. A knot sequence $\boldsymbol{t} = \{t_0, t_1, \ldots, t_{m-1}, t_m\}$ is defined as a non-decreasing and finite sequence of real numbers that define the partitioning of the domain [a,b], i.e. $a = t_0 \le t_1 \le \cdots \le t_{m-1} \le t_m = b$, such that $S(x)$ is a polynomial on any interval $[t_{i}, t_{i+1}]$. The number of B-spline basis functions $k$ relates to the degree $\rho$ and number of knots $m+1$ by $\label{start} k = m-\rho$. B-spline basis functions of arbitrary degree $\rho$ can be recursively constructed by means of the Cox-de Boor formula \cite{practical}:
\begin{dmath}
\label{eq:4.5}
B_{i,0}(x) = \left\{
\begin{matrix} 
1 & \mathrm{if} \quad t_{i-1} \leq x < t_{i} \\
0 & \mathrm{otherwise} 
\end{matrix}
\right. ,\\
{B_{i,\rho+1}(x) = \frac{x - t_{i-1}}{t_{i+\rho-1} - t_{i-1}} B_{i,\rho}(x) + \frac{t_{i+\rho} - x}{t_{i+\rho} - t_{i}} B_{i+1,\rho}(x)} .
\end{dmath}
If the knots are equidistantly distributed over the domain of the spline, the spline is called uniform. If the uniform knot sequence is also a subset of $\mathbb{Z}$, i.e. a sequence of integers, the spline is referred to as a cardinal spline \cite{schoenberg1973cardinal}. In this article, all knot sequences will be considered uniform, as they allow for efficient evaluation of $\boldsymbol{b}$ using a matrix expression \cite{qin2000general} instead of \eqref{eq:4.5}.
Figure \ref{fig:cardinal} illustrates B-splines of degree 1 to 3 on the cardinal knot sequence $\boldsymbol{t} = \{0, 1, 2, 3, 4, 5\}$. The dashed purple lines represent the sum of the basis functions. The shape of a B-spline curve $S(x)$ is only fully adjustable within its natural domain $\mathcal{D}_n=[t_{\rho}, t_{m-\rho}]$, because the sum of the B-spline basis functions at any point within this domain equals one. It is desirable to have full control over the shape of the B-spline curve over the whole range of data samples. The knot sequences in this article will be chosen such that $\mathcal{D}_n$ coincides with the unit interval $[0, 1]$. 

\subsection{Multivariate B-splines}

B-splines generalize to multiple input dimensions through tensor products of univariate basis functions. One can construct a $d$-dimensional spline $S$ as a linear combination of multivariate B-splines:
\begin{align}\label{multivariate}
&S(x_1,x_2,\ldots,x_d) \nonumber \\
&= \sum_{i_1=1}^{k_1} \sum_{i_2=1}^{k_2} \cdots \sum_{i_d=1}^{k_d} 
 B_{i_1}(x_1) B_{i_2}(x_2) \cdots B_{i_d}(x_d) \mathcal{W}_{i_1 i_2 \cdots i_d} \nonumber\\ 
&= \left< \mathcal{B}, \mathcal{W} \right>.
\end{align}
For notational convenience, we omitted the degrees $\rho$. The B-spline tensor $\mathcal{B}$ contains the multivariate basis functions and is defined as: 
\begin{align*} 
\mathcal{B} &=
\boldsymbol{b}^{(1)} \circ  \boldsymbol{b}^{(2)} \circ  \cdots \circ  \boldsymbol{b}^{(d)},
\end{align*}
where $\boldsymbol{b}^{(p)}$ is the univariate basis vector of the $p$-th input variable, i.e.
\begin{dmath}
\boldsymbol{b}^{(p)} = 
\begin{bmatrix} 
B_{1,\rho}(x_p) & B_{2,\rho}(x_p) & \cdots & B_{k_p,\rho}(x_p)
\end{bmatrix} ^T .%\condition{$p \in {1,2,\ldots,d}$}
\end{dmath}
We will assume equal knots and degree for each dimension, hence $k_p = k$, $\, \forall p$. The representation of B-spline surfaces in \eqref{multivariate} is severely limited by the exponential increase in the number of basis functions and weights, $\mathcal{O}((k)^d)$. 

\section{Tensor Network B-splines}

For our purposes, the input variables $x_p$ are the lagged inputs and outputs of \eqref{system}. For a large number of lags or inputs, it can therefore quickly become computationally infeasible to store or operate on the tensors $\mathcal{B}$ and $\mathcal{W}$. Using tensor network theory, the multivariate B-spline surface can be represented in a low-parametric format. In this section, we derive the TNBS model and use it to approximate the function $f$ in \eqref{system} from observed input and output data.
\subsection{Model structure}
We illustrate the model structure using a three-dimensional Tensor Network B-spline surface as an example, which is derived as follows:
\begin{figure}
    \centering
    \includegraphics[scale=0.25]{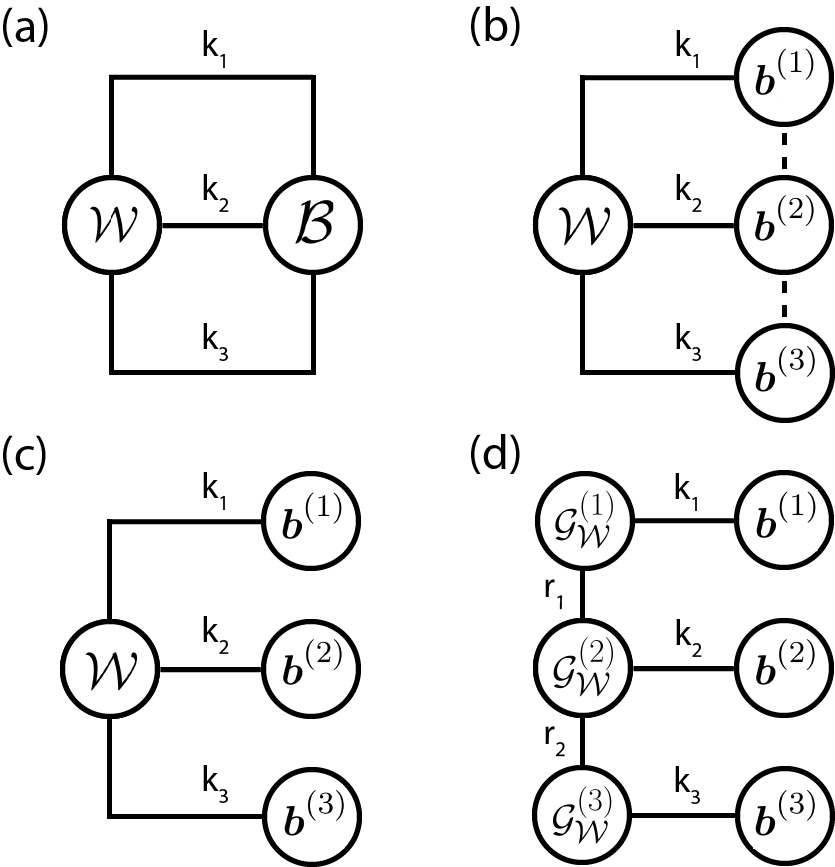}
    \caption{Derivation of the Tensor Network B-splines model of order 3.}
    \label{fig:ttbs}
\end{figure} 
\begin{align}
S(&x_1,x_2,x_3)\nonumber \\
&= \left< \mathcal{B}, \mathcal{W} \right> \label{seven}\\
&=  \mathcal{W} \times_{1}^1 \boldsymbol{b}^{(1)} \times_{2}^1 \boldsymbol{b}^{(2)}  \times_{3}^1 \boldsymbol{b}^{(3)} \label{eight}\\
&=  (\mathcal{G}_{\mathcal{W}}^{(1)} \times_{2}^1 \boldsymbol{b}^{(1)})
 (\mathcal{G}_{\mathcal{W}}^{(1)} \times_{2}^1 \boldsymbol{b}^{(2)})
  (\mathcal{G}_{\mathcal{W}}^{(1)} \times_{2}^1 \boldsymbol{b}^{(3)}) \label{nine}.
\end{align}
Figure \ref{fig:ttbs} will be used as a visual reference to walk through these equations. Given the weight tensor $\mathcal{W} \in \mathbb{R}^{k_1\times k_2 \times k_3}$ and B-spline tensor $\mathcal{B} \in \mathbb{R}^{k_1\times k_2 \times k_3}$ in \eqref{seven}, their inner product is equal to the contraction over all pairs of indexes, as seen in Figure \ref{fig:ttbs}a. As $\mathcal{B}$ is a rank one tensor, it can be decomposed into the outer product of three B-spline vectors $\boldsymbol{b}^{(p)}$ (Figure \ref{fig:ttbs}b). The outer product operation is a special case of contraction where the contracted indexes have singleton dimensions. Singleton contractions that close a loop in a tensor network are redundant, and hence omitted in Figure \ref{fig:ttbs}c. Now $S(x_1,x_2,x_3)$ in \eqref{eight} is simply the contraction of $\mathcal{W}$ with the B-spline basis vectors. Finally, $\mathcal{W}$ is decomposed into a tensor train in Figure \ref{fig:ttbs}d. A point ${(x_1, x_2,x_3)}$ on the TNBS surface in  \eqref{nine} is evaluated by constructing the B-spline vectors $\boldsymbol{b}^{(p)}$, contracting them with the corresponding tensor train cores and finally multiplying the sequence of resulting matrices. Due to the constraints, $r_{0} = r_{d} = 1$, this results in a scalar output. Extending the TNBS model to $l$ outputs can be realized by removing one of these constraints, e.g. $r_{0} = l$. In general, a $d$-dimensional TNBS surface is represented by:
\begin{align}\label{ttbs}
S(x_1,x_2,\ldots,x_d) 
= \prod_{p=1}^d (\mathcal{G}_{\mathcal{W}}^{(p)} \times_{2}^1 \boldsymbol{b}^{(p)}).
\end{align}

\subsection{Identification algorithm} 
\begin{figure}
    \centering
    \includegraphics[scale=0.25]{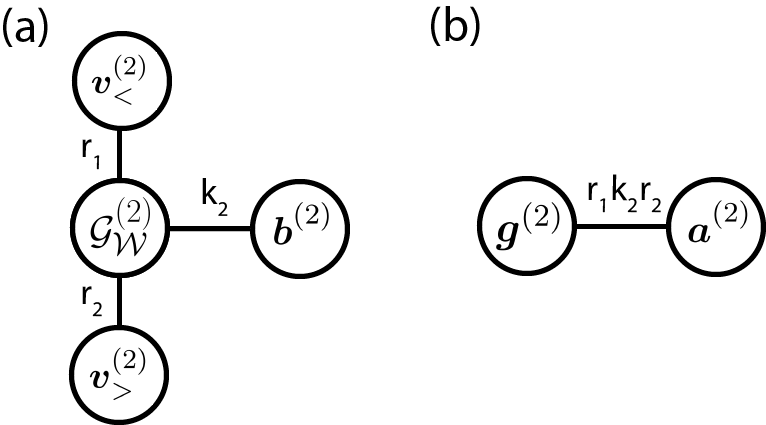}
    \caption{The tensor network written as a vector inner product.}
    \label{linear}
\end{figure}

\begin{figure}
    \centering
    \includegraphics[scale=0.25]{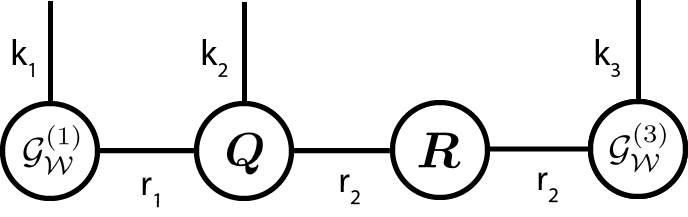}
    \caption{QR decomposition of the second core during a left to right sweep.}
    \label{fig:qr}
\end{figure}

We illustrate, without loss of generality, the proposed identification algorithm by means of the following example. Suppose we have the following NARX system model:
\begin{equation}\label{systema}
     y_n = f(u_{n},y_{n-1}, u_{n-1}) + \varepsilon_n.
\end{equation}
We want to identify this model from a set of observed input and output data  $\{(y_{n}, u_{n})\}_{n=1}^N$. We approximate the function $f$ with the three-dimensional TNBS from Figure \ref{fig:ttbs}d, by minimizing the least-squared cost function:
\begin{align}\label{error}
\min_{\mathcal{W}} \,\,\,& \| \boldsymbol{y} - \boldsymbol{s} \|^2_2\\ \nonumber
\text{s.t.}  \, \text{TT-rank}(\mathcal{W}) &= (r_1,r_2),
\end{align}
where 
\begin{equation*}
\boldsymbol{y} = 
\begin{bmatrix}
y_2 \\ y_3 \\ \vdots \\ y_N
\end{bmatrix}, \qquad
\boldsymbol{s} = 
\begin{bmatrix}
 f( u_{2}, y_{1}, u_{1}) \\
 f( u_{3}, y_{2}, u_{2}) \\
 \vdots \\ 
 f(u_{N},y_{N-1}, u_{N-1})
\end{bmatrix}.
\end{equation*}
To solve \eqref{error} directly for the TT-cores, we use the alternating linear scheme (ALS) \cite{holtz2012alternating}. The TT-ranks are chosen beforehand and the TT-cores are initialized randomly. ALS then iteratively optimizes one tensor core at a time while holding the others fixed. Optimizing one core is equal to solving a small linear subsystem. Suppose we wish to update the second core from Figure \ref{fig:ttbs}d. The idea is to contract everything in the network up until the nodes adjacent to $\mathcal{G}_{\mathcal{W}}^{(2)}$ (Figure \ref{linear}a), whereupon \eqref{kim} is used to rewrite the network as an inner product of two vectors (Figure \ref{linear}b):
\begin{dmath}\label{explain}
y_n =  \mathcal{G}_{\mathcal{W}}^{(2)} \times_{1}^1 \boldsymbol{v}_{<}^{(2)} \times_{2}^1 \boldsymbol{b}^{(2)}  \times_{3}^2 \boldsymbol{v}_{>}^{(2)}\\
= \mathcal{G}_{\mathcal{W}}^{(2)} \times_{1}^2 \boldsymbol{v}_{<}^{(2)T} \times_{2}^2 \boldsymbol{b}^{(2)T}  \times_{3}^2 \boldsymbol{v}_{>}^{(2)}
= \left( \boldsymbol{v}_{>}^{(2)T} \otimes \boldsymbol{b}^{(2)T} \otimes \boldsymbol{v}_{<}^{(2)} \right) \boldsymbol{vec}\left(\mathcal{G}_{\mathcal{W}}^{(2)}\right)\\
= \boldsymbol{a}^{(2)T} \boldsymbol{g}^{(2)} .
\end{dmath}
More generally, rewriting \eqref{ttbs} for the $n$-th data sample as a linear function of the elements of the $p$-th core gives:
\begin{dmath}\label{lin}
y_n = \left( \boldsymbol{v}_{>,n}^{(p)T} \otimes \boldsymbol{b}_n^{(p)T} \otimes \boldsymbol{v}_{<,n}^{(p)} \right) \boldsymbol{vec}\left(\mathcal{G}_{\mathcal{W}}^{(p)}\right),
\end{dmath}
where
\begin{align*}
\boldsymbol{v}_{<,n}^{(p)} &= 
\prod_{j=1}^{p-1} (\mathcal{G}_{\mathcal{W}}^{(j)} \times_{2}^1 \boldsymbol{b}_n^{(j)}) \in \mathbb{R}^{1 \times r_{p-1}}\\
\boldsymbol{v}_{>,n}^{(p)} &= 
\prod_{j=p+1}^{d} (\mathcal{G}_{\mathcal{W}}^{(j)} \times_{2}^1 \boldsymbol{b}_n^{(j)}) \in \mathbb{R}^{r_{p}}
\end{align*}
for $2 \leq p \leq d-1$, and $
\boldsymbol{v}_{<,n}^{(1)} = \boldsymbol{v}_{>,n}^{(d)} = 1$.
Computing \eqref{lin} for all $N$ data samples results in a system of linear equations. The subproblem for updating the $p$-th core thus becomes:
\begin{equation}\label{suberror}
\min_{\boldsymbol{g}^{(p)}} \,\,\, \left\| \boldsymbol{y} - \boldsymbol{A}^{(p)} \boldsymbol{g}^{(p)} \right\|^2_2,
\end{equation}
where 
\begin{equation}\label{AA}
\boldsymbol{A}^{(p)} = 
\begin{bmatrix}
 \boldsymbol{v}_{>,1}^{(p)T} \otimes \boldsymbol{b}_1^{(p)T} \otimes \boldsymbol{v}_{<,1}^{(p)}\\
\boldsymbol{v}_{>,2}^{(p)T} \otimes \boldsymbol{b}_2^{(p)T} \otimes \boldsymbol{v}_{<,2}^{(p)}\\ 
 \vdots\\
 \boldsymbol{v}_{>,N}^{(p)T} \otimes \boldsymbol{b}_N^{(p)T} \otimes \boldsymbol{v}_{<,N}^{(p)} 
\end{bmatrix}, \hspace{3mm} \boldsymbol{g}^{(p)}  = \boldsymbol{vec}\left(\mathcal{G}_{\mathcal{W}}^{(p)}\right).
\end{equation}
The optimum is found by solving the normal equation:
\begin{equation}\label{normala}
\left(\boldsymbol{A}^{(p)T}\boldsymbol{A}^{(p)}\right) \boldsymbol{g}^{(p)} = \boldsymbol{A}^{(p)T}\boldsymbol{y}.
\end{equation}
Reshaping $\boldsymbol{g}^{(p)}$ back into a third-order tensor results in the updated core $\mathcal{G}_{\mathcal{W}}^{(p)}$. The ALS algorithm sweeps back and forth, iterating from the first to the last core and back, until convergence. At each iteration, \eqref{normala} is solved for $\boldsymbol{g}^{(p)}$. Numerical stability is ensured by keeping the tensor train in site-p-mixed-canonical form through an additional orthogonalization step. To illustrate, consider again the TNBS in Figure \ref{fig:ttbs}d. Assume that we are iterating from left to right and the tensor train is in site-2-mixed-canonical form. After solving $\boldsymbol{g}^{(p)}$ it is reshaped into a matrix $\boldsymbol{G}^{(p)} \in \mathbb{R}^{r_{p-1} k_p\times r_{p}}$, which is then decomposed through a QR decomposition. The tensor network is now in the form of Figure \ref{fig:qr}.
Finally, $Q$ is reshaped back into a third-order left-orthogonal tensor $\boldsymbol{G}^{(2)}$ and $R$ is contracted with the next core. The tensor train is now in site-3-mixed-canonical form, and the next iteration starts. More details about the orthogonalization step are given in \cite{holtz2012alternating}. The optimization with ALS converges monotonously, so a possible stopping criterion is:
\begin{equation}
\left\| J_h^{(1)} -  J_{h+1}^{(1)} \right\|_2 \, \leq \epsilon,
\end{equation}
where $J_h^{(1)}$ is the cost of the objective function in \eqref{suberror} during the first core update of the $h$-th sweep. A modified version of ALS method, MALS \cite{holtz2012alternating}, updates two cores simultaneously and is computationally more expensive, but is able to adaptively determine the optimal TT-ranks for a specified accuracy. Another adaptive method is the tensor network Kalman filter \cite{batselier2017kalman}, which can be used for online optimization of the cores. 

\subsection{Regularization}

\begin{figure}
    \centering
    \includegraphics[scale=0.25]{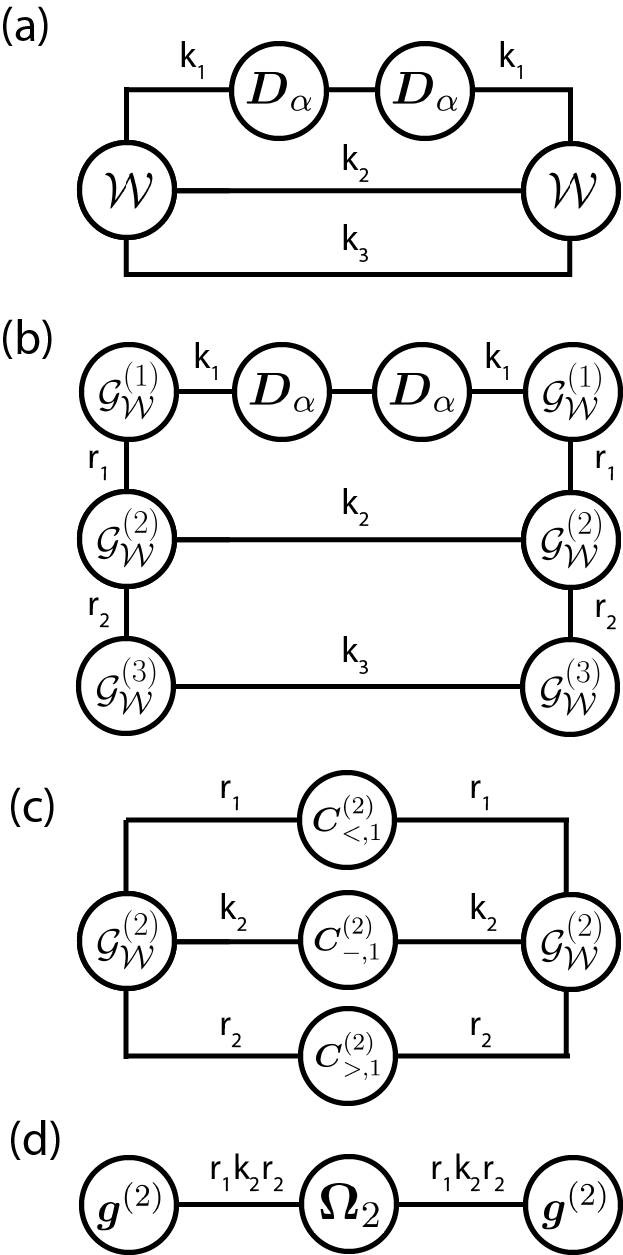}
    \caption{Derivation of the Tensor Network P-spline penalty.}
    \label{penal}
\end{figure}

In addition to decreasing computational burden, the TT-rank constraints serve as a regularization mechanism. This regularization is however insufficient for high-dimensional B-splines, as the volume of the domain of the TNBS increases exponentially. The available estimation data becomes sparse and scattered, which can lead to an ill-posed optimization problem. B-spline curves inherently possess the ability to regularize by adjustment of their degree or knot placement. The choice of knots has been a subject of much research \cite{eubank1999nonparametric}, but due to lack of an attractive all-purpose scheme, we opt for a non-parametric approach known as P-splines \cite{eilers1996flexible}. P-splines induce smoothness by combining uniform B-splines with a discrete penalty placed on the $\alpha$-th difference between adjacent weights. For univariate splines, the following penalty function is added to the cost function: 
\begin{equation}\label{pspline}
R(\boldsymbol{w}) =  \,  \| \boldsymbol{D}_\alpha \boldsymbol{w} \|^2_2 .
\end{equation}
The matrix $\boldsymbol{D}_\alpha \in \mathcal{R}^{(k+1-\alpha)\times (k+1)} $ is the $\alpha$-th order difference matrix such that $\boldsymbol{D}_\alpha \boldsymbol{w} = \Delta^\alpha\boldsymbol{w}$ results in a vector of $\alpha$-th order differences of $\boldsymbol{w}$. This matrix can be constructed by using the difference operator $\alpha$ times consecutively on the identity matrix. For $\alpha = 0$ this is equal to Tikhonov regularization and for $\alpha = 1$ we get Total Variation regularization. For example, given are a weight vector and the first-order difference matrix:
$$\boldsymbol{w} =
\begin{bmatrix}
    w_{1} \\
    w_{2} \\
    w_{3} \\
\end{bmatrix}, \qquad 
\boldsymbol{D}_1 =
\begin{bmatrix}
    1      & -1 & 0 \\
    0       & 1 & -1 \\
\end{bmatrix}.
$$
The penalty term then equals:
$$\|\boldsymbol{D}_1 \boldsymbol{w}\|_2^2 = 
(\boldsymbol{D}_1 \boldsymbol{w})^T (\boldsymbol{D}_1 \boldsymbol{w})
$$
$$ =
\begin{bmatrix}
    (w_{1} - w_{2}) &
    (w_{2} - w_{3})
\end{bmatrix}
\begin{bmatrix}
    (w_{1} - w_{2}) \\
    (w_{2} - w_{3}) \\
\end{bmatrix}
$$
$$
=    (w_{1} - w_{2})^2 +
    (w_{2} - w_{3})^2 .$$
We wish to extend the penalty in \eqref{pspline} to the TNBS format. Without loss of generality, Figure \ref{penal} visualizes the necessary steps in graphical notation for a three-dimensional B-spline surface. In the multivariate case, the differences in adjacent weights in the weight tensor $\mathcal{W}$ have to be penalized along each dimension individually. This is done by contracting the second index of the difference matrix $\boldsymbol{D}_\alpha$ with the dimension of the weight tensor $\mathcal{W}$ along which the penalty is applied, then taking the norm of the result. For a B-spline curve with $d$ inputs, the penalty on the $\alpha$-th order differences along the $j$-th dimension is given by: 
\begin{align}
R(\mathcal{W}) &= \nonumber\|\mathcal{W} \, \times_{j}^{2} \,\boldsymbol{D}_{\alpha}\|_2^2 \\
&= \left<( \mathcal{W} \, \times_{j}^{2} \,\boldsymbol{D}_{\alpha}) \,, \,\, (\mathcal{W} \, \times_{j}^{2} \,\boldsymbol{D}_{\alpha}) \right>.
\end{align}
This is illustrated in \ref{penal}a, where the penalty is applied along the first dimension, e.g. $j=1$. 
Decomposing $\mathcal{W}$ into a tensor train results in the network depicted in \ref{penal}b. To write this penalty again as a linear function of the $p$-th core, we contract everything in the network except these cores (Figure \ref{penal}c). In this example, $\boldsymbol{C}_{>,1}^{(2)}$ and $\boldsymbol{C}_{-,1}^{(2)}$ are simply identity matrices. Then, using \eqref{kim}, the penalty function can be rewritten in the form of Figure \ref{penal}d:
\begin{align}
&R\left(\mathcal{G}_{\mathcal{W}}^{(p)}\right) = \boldsymbol{g}^{(p)T} \boldsymbol{\Omega}_{j}^{(p)} \boldsymbol{g}^{(p)},
\end{align}
where
\begin{equation}\label{omega}
    \boldsymbol{\Omega}_{j}^{(p)}= \left( \boldsymbol{C}_{>,j}^{(p)} \otimes \boldsymbol{C}_{-,j}^{(p)} \otimes\boldsymbol{C}_{<,j}^{(p)}\right) .
\end{equation}
The matrix $\boldsymbol{\Omega}_{j}^{(p)}$ in \eqref{omega} is constructed for every dimension $j$. Due to the site-p-mixed-canonical form of the tensor train, the contraction of two out of the three matrices $\boldsymbol{C}_{>,j}^{(p)}$, $\boldsymbol{C}_{-,j}^{(p)}$ and $\boldsymbol{C}_{<,j}^{(p)}$ result in identity matrices. This knowledge can be utilized for efficient implementation. Adding the penalties to the cost function results in the following regularized optimization problem:
\begin{align}\label{psplines}
\min_{\mathcal{W}}   \,\,\, \| \boldsymbol{y} - \boldsymbol{s} \|^2_2 + 
\sum_{j=1}^d \lambda_j \,  
\|\mathcal{W} \, \times_{j}^{2} \,\boldsymbol{D}_{\alpha}\|_2^2\\ \nonumber
\text{s.t. }\,  \text{TT-rank}(\mathcal{W}) = (r_1,r_2,\ldots,r_{d-1})  .
\end{align}
The smoothing parameter $\lambda_j \geq 0$ controls the penalization of the roughness along dimension $j$.
The subproblem for updating the $p$-th core becomes:
\begin{equation}\label{psplineproblem}
\min_{\boldsymbol{g}^{(p)}} \,\,\, \| \boldsymbol{y} - \boldsymbol{A}^{(p)}\boldsymbol{g}^{(p)} \|^2_2 + 
\sum_{j=1}^d \lambda_j \,  \boldsymbol{g}^{(p)T} \boldsymbol{\Omega}^{(p)}_j \boldsymbol{g}^{(p)} .
\end{equation}
The normal equation is then:
\begin{equation}\label{penala}
\left(\boldsymbol{A}^{(p)T}\boldsymbol{A}^{(p)} + \sum_{j=1}^d \lambda_j \, \boldsymbol{\Omega}^{(p)}_j \right) \boldsymbol{g}^{(p)} = \boldsymbol{A}^{(p)T}\boldsymbol{y}.
\end{equation}
The whole procedure of identifying a TNBS model from measured data is summarized as pseudo-code in Algorithm \ref{algo}.

\begin{algorithm}[h]
\caption{TNBS-NARX identification}\label{algo}
 \begin{algorithmic}[1] 
\Statex \textbf{Input:} Data $\{(y_{n}, u_{n})\}_{n=1}^N$, TT-ranks $\{r_p\}_{p=1}^d$, number of knots $m$, degree $\rho$, regularization
\Statex  parameters $\{\lambda_j\}_{j=1}^d$
\Statex \textbf{Output:} TT-cores $ \left\{\mathcal{G}_{\mathcal{W}}^{(p)}\right\}{}_{p=1}^d$
\State Initialize random TT-cores
\State Construct $\left\{\left\{\boldsymbol{b}_n^{(p)}\right\}{}_{n=1}^N\right\}{}_{p=1}^d$ from data
\While {stopping criteria not satisfied}
  \For {$p=1,2,\ldots,d-1$}
    \State Construct $\boldsymbol{A}^{(p)}$ \eqref{AA}  and $\left\{\boldsymbol{\Omega}^{(p)}_j\right\}{}_{j=1}^d$ \eqref{omega}
    \State $\boldsymbol{g}^{(p)} \gets$ Solve \eqref{penala}
    \State $\mathcal{G}_{\mathcal{W}}^{(p)} \gets$ Orthogonalize and reshape $\boldsymbol{g}^{(p)}$
  \EndFor
    \For {$p=d,d-1,\ldots,2$}
\State Repeat the above
  \EndFor
\EndWhile
\end{algorithmic}
\end{algorithm}
Table \ref{complex} summarizes relevant computational complexities concerning the TNBS-NARX method. While the complexities scale only linearly in the dimensions, it is important to realize that high TT-ranks easily degrade the performance of optimization of the cores. There is, therefore, a tradeoff between accuracy and speed. The number of data samples $N$ also appears linearly in the complexities but may become a bottleneck for large datasets. A modification for this scenario is to use a small random batch of the data when updating $\boldsymbol{g}^{(p)}$. This can speed up estimation time without significant loss of accuracy.
\begin{table}[h]
\caption{\label{complex}Computational complexities of significant operations}
\begin{center}
\begin{tabular}{ c c }
\hline
 Operation & Complexity\\ 
\hline
 Construct $\left\{\boldsymbol{b}_n^{(p)}\right\}{}_{n=1}^N$ & $O\left(Nn^2\right)$ \\
 Construct $\left\{\boldsymbol{\Omega}^{(p)}_j\right\}{}_{j=1}^d$ & $O\left((d+(m-\rho)^4)r^4\right)$ \\
 Construct $\boldsymbol{A}^{(p)}$ & $O\left(N(m-\rho)r^2\right)$ \\
 Solve $\boldsymbol{g}^{(p)}$ & $O\left(N(m-\rho)^2r^4+(m-\rho)^3r^6\right)$\\  
 Evaluate $f$  & $O\left((\rho^2+(m-\rho)r^2)d\right)$ \\
 \hline
\end{tabular}
\end{center}
\end{table}

\section{Experiments}
In this section, we demonstrate the proposed system identification method. The algorithm is implemented in MATLAB and executed on a personal computer with a 4.2 GHz Intel Core i5-7600K processor and 16 GB of random access memory (RAM). An open-source MATLAB implementation can be found at \url{https://github.com/Ridvanz/Tensor-Network-B-splines}.

\subsection{Synthetic dataset}

\begin{figure}[t]\label{plot}
    \begin{center}
    \includegraphics[width=8.4cm]{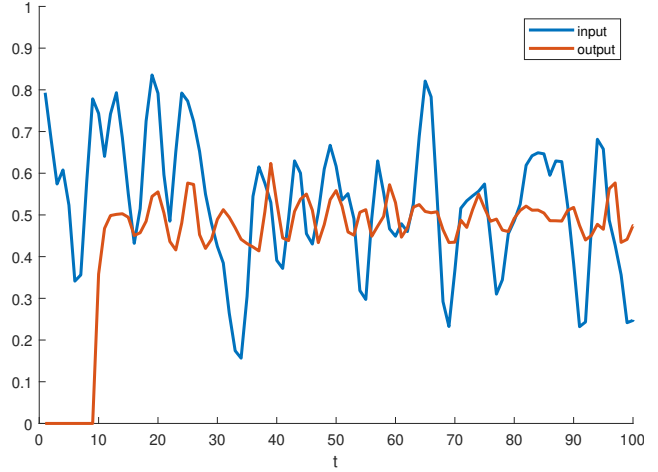}    % The printed column  
    \caption{Input and output signals of the synthetic dataset.}  % width is 8.4 cm.
    \label{inout}                                 % Size the figures 
    \end{center}                                 % accordingly.
\end{figure}

First, we validate the proposed methods through the identification of an artificial nonlinear dynamical system that is exactly representable in the TNBS-NARX format. The lagged inputs and outputs are chosen as $u_{n-\mu}$ and $y_{n-\mu}$ respectively, where $\mu \in (1, 2, 3, 4)$, such that the system equation is of the following form:
\begin{equation}
    y_n = f(y_{n-1}, y_{n-2}, y_{n-3}, y_{n-4}, u_{n-1}, u_{n-2}, u_{n-3}, u_{n-4})\label{narxx}
\end{equation}
The nonlinear mapping $f$ is modeled as an 8-dimensional TNBS. We choose the degree of the B-splines $\rho=2$ and the number of knots per dimension $m=6$. A random weight tensor $\mathcal{W}$ of size ${(m-\rho)^d = 4^8}$ is generated of which the elements equal either $w_{min}=-4$ or $w_{max}=5$ with equal probability. The generated tensor is decomposed using the TT-SVD algorithm, truncating the TT-ranks to a value of $5$ uniformly. The resulting tensor train represents the true weights of our nonlinear system. For the input signal $\boldsymbol{u}$ we generate a random sequence of length $3000$, with values uniformly distributed in the range $[0,1]$. This sequence is smoothed with a Gaussian window of size 5 to dampen higher frequencies. We initialize the output signal $\boldsymbol{y}$ with $4$ zeros and recursively evaluate the next output with \eqref{narxx}, until we have a signal of length 3000. The first 200 samples of the input (blue) and output (red) signals are plotted in in Figure \ref{plot}. The signals are split in an identification set of $2000$ samples and a test set of $1000$ samples.

\begin{figure}
    \centering
    \includegraphics[width=8.4cm]{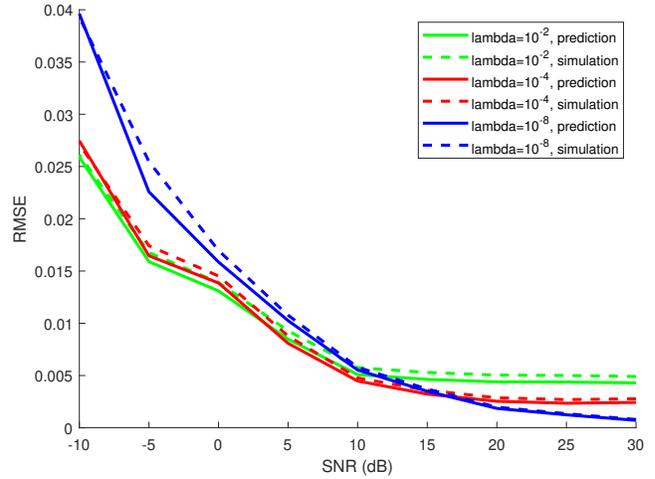}
    \caption{Prediction and simulation performance on synthetic test set.}
    \label{synthtest}
\end{figure}

We test the performance of our TNBS-NARX identification algorithm with different levels of Gaussian white noise on the estimation data. Noise is only added to the output signal. The variances for the white noise signals are chosen based on the desired signal to noise ratios SNR. The signal powers are determined after subtracting their means. For simplicity, we penalize the second difference ($\alpha = 2$) of the weights equally for each dimension, e.g. $\lambda_p = \lambda$, $\, \forall p$. The experiment is run using three different values for lambda. All other model parameters are set to the true values of the synthetic model. The TT-cores are estimated using Algorithm \ref{algo}. For consistency, we simply choose a max number (16) of sweeps as stopping criteria. The root mean squared error (RMSE) is used as the performance metric to evaluate the accuracy on the test set for both prediction and simulation. 
$$e_{RMSE} = \sqrt{\frac{1}{N} \sum_{i=1}^N  (y_i - \hat{y_i})^2 }$$
Figure \ref{synthtest} plots the RMSE of the different experiments as a function of the SNR in dB. The prediction errors are consistently lower than the simulation errors. The effect of the regularization is in line with expectations, i.e. for increasing SNR values, more regularization is needed to avoid overfitting to noise, so larger penalties give better performance. Overall, the TNBS is able to identify the system accurately, even for relatively noisy estimation data.

\subsection{Cascaded tanks dataset}
The cascaded tanks system is a benchmark dataset for nonlinear system identification.  A detailed description of the system and the data is given in \cite{schoukens2016cascaded}. The system consists of two tanks, a water reservoir and a pump. The water in the reservoir is pumped in the upper tank, from which it flows to the lower tank through a small opening and then back into the reservoir. The system input $u_n$ is the pump voltage and the system output $y_n$ is the water level of the lower tank. If too much water is pumped into the upper tank it overflows, causing a hard saturation nonlinearity in the system dynamics. The input signals are low-frequency multisine signals. Both the estimation and test set have a length of $N=1024$ samples. The major challenges of this benchmark are the hard saturation nonlinearity and the relatively small size of the estimation set. The performance metric used is again RMSE.

\begin{figure}
    \centering
    \includegraphics[width=8.4cm]{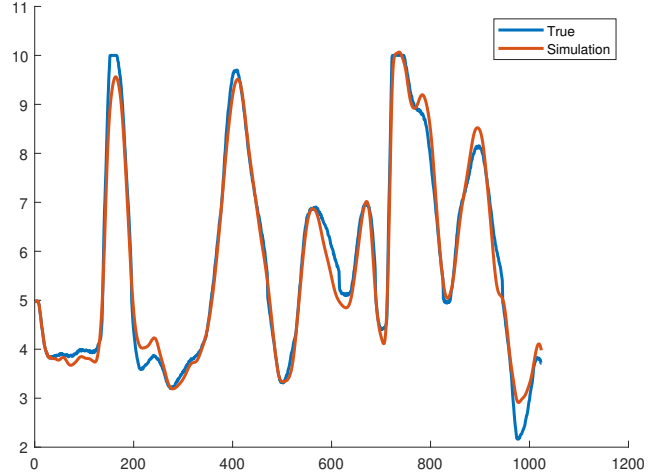}
    \caption{Simulation on cascaded tanks dataset.}
    \label{fig:simulatecas}
\end{figure}

The original data is first normalized to the interval [0,1]. Both input and output lags are chosen as $u_{n-\mu}$ and $y_{n-\mu}$ respectively, where $\mu \in \{1, 2, 3, 4, 8, 12, 16, 32\}$. The large lags are included to capture the relevant slow system dynamics. We choose the degree of the B-splines $\rho=3$ and the number of knots $m=7$. We penalize the first-order difference only, i.e. $\alpha = 1$, and set the TT-ranks to $8$ uniformly. We choose $\lambda$ through 3-fold cross-validation on the estimation set. A total of 12 sweeps are performed in the optimization with Algorithm \ref{algo}. After tuning lambda, the full identification set is used to identify the final model, which takes about $4$ seconds. Using TNBS, the number of weights to represent the $16$-dimensional B-spline surface is reduced from approximately $4.3 \times 10^{9}$ to $3648$. The performance on prediction and simulation are listed and compared in table \ref{results}. To the best of our knowledge, the algorithm slightly outperforms the current state-of-the-art results on both prediction and simulation. Figure \ref{fig:simulatecas} shows the true and simulated output on the test set. It is apparent that the TNBS-NARX model was able to accurately capture the nonlinear system dynamics with relatively sparse estimation data. 

\begin{table}
\caption{\label{results}Comparison of methods on Cascaded tanks benchmark}
\begin{center}
\begin{tabular}[b]{ c c c }
\hline
 Method & Prediction & Simulation \\ 
\hline
 LTI \cite{schouk}   & 0.056 &  0.588 \\
 Volterra FB \cite{schouk}  & 0.049 &  0.397 \\
 Flexible SS \cite{svensson2017flexible} & - &  0.45 \\
 NOMAD \cite{brunot2017continuous} & - &   0.376 \\
 PWARX \cite{mattsson2018identification} & - &   0.350 \\  
 Sparse Bay. DNN \cite{zhou2019sparse} &  0.0472 &   0.344 \\
 \hline
  TNBS-NARX & 0.0461 & 0.3018 \\ %/ 0.0969
 \hline
\end{tabular}
\end{center}
\end{table}
\section{Conclusions}
This article presents a new algorithm for nonlinear system identification using a NARX model of which the nonlinear mapping is approximated using the introduced Tensor Network B-splines. Tensor Network theory enables to work with B-spline surfaces directly in a high-dimensional feature space, allowing the identification of NARX systems with a large number of lags and inputs. The identification algorithm is guaranteed to monotonically converge and numerical stability is ensured through orthogonality of the TT-cores. The efficiency and accuracy of the algorithm is demonstrated through numerical experiments on SISO nonlinear systems. Extension of TNBS-NARX to multiple inputs is straightforward through the addition of input variables. Multiple outputs can be realized efficiently by adding an index to one of the TT-cores, as done in \cite{batselier2017tensor}. Future work includes the implementation of an online optimization scheme, as an alternative to ALS, and the development of control strategies for identified TNBS-NARX systems.

\bibliographystyle{unsrt}        % Include this if you use bibtex 
% % unsrt ipv plain
\bibliography{autosam}           % and a bib file to produce the 
                                 % bibliography (preferred). The
                                 % correct style is generated by
                                 % Elsevier at the time of printing.

%\begin{thebibliography}{99}     % Otherwise use the  
                                 % thebibliography environment.
                                 % Insert the full references here.
                                 % See a recent issue of Automatica 
                                 % for the style.
%  \bibitem[Heritage, 1992]{Heritage:92}
%     (1992) {\it The American Heritage. 
%     Dictionary of the American Language.}
%     Houghton Mifflin Company.
%  \bibitem[Able, 1956]{Abl:56}
%     B.~C.~Able (1956). Nucleic acid content of macroscope. 
%     {\it Nature 2}, 7--9. 
%  \bibitem[Able {\em et al.}, 1954]{AbTaRu:54}   
%     B.~C. Able, R.~A. Tagg, and M.~Rush (1954).
%     Enzyme-catalyzed cellular transanimations.
%     In A.~F.~Round, editor, 
%     {\it Advances in Enzymology Vol. 2} (125--247). 
%     New York, Academic Press.
%  \bibitem[R.~Keohane, 1958]{Keo:58}
%     R.~Keohane (1958).
%     {\it Power and Interdependence: 
%     World Politics in Transition.}
%     Boston, Little, Brown \& Co.
%  \bibitem[Powers, 1985]{Pow:85}
%     T.~Powers (1985).
%     Is there a way out?
%     {\it Harpers, June 1985}, 35--47.

%\end{thebibliography}

\end{document}